\documentclass[12pt,preprint]{aastex}
\usepackage{verbatim}
\shorttitle{The Two Modes of Gas Giant Planet Formation}
\shortauthors{Boley, A.~C.}

\begin{document}
\title{The Two Modes of Gas Giant Planet Formation}

\author{Aaron C.\ Boley}

\affil{Institute for Theoretical Physics, University of Zurich, Winterthurerstrasse 190,
Zurich, CH-8057, Switzerland; aaron.boley@gmail.com}

\begin{abstract}
I argue for two modes of gas giant planet formation and discuss the conditions under which each mode operates.  Gas giant planets at disk radii $r>100$ AU are likely to form in situ by disk instability, while core accretion plus gas capture remains the dominant formation mechanism for $r<100$ AU. During the mass accretion phase, mass loading can push disks toward fragmentation conditions at large $r$.  Massive, extended disks can fragment into clumps of a few to tens of Jupiter masses.  This is confirmed by radiation hydrodynamics simulations.   The two modes of gas giant formation should lead to a bimodal distribution of gas giant semi-major axes.  Because core accretion is expected to be less efficient in low-metallicity systems, the ratio of gas giants at large $r$  to planets at small $r$ should increase with decreasing metallicity.
\end{abstract}


\section{Introduction}

%Is core accretion plus gas capture  (e.g., Pollack et al.~1996) or direct formation by disk instability (Cameron 1978; Boss 1997) the principal formation mechanism for gas giant planets?

Both core accretion plus gas capture (e.g., Pollack et al.~1996) and direct formation by disk instability (Cameron 1978; Boss 1997) can form gas giant planets in principle.
 In order for gravitational instabilities (GIs) to form gas giants directly,
 the Toomre (1964) $Q=c_s\kappa/\pi G\Sigma$, where  $c_s$ is the sound
speed, $\kappa$ is the epicyclic frequency, and $\Sigma$ is the surface
density, must approach unity {\it and} the local cooling time must be
$\lesssim$ the local orbit period (Gammie 2001; Rice et al.~2005; Rafikov
2005, 2007).  An isothermal, low-$Q$ is strongly susceptible to fragmentation
(e.g., Tomley et al.~1994; Boss 1998; Nelson et al.~1998; Mayer et al.~2002; Pickett et al.~2003).

Analytical work (e.g., Rafikov 2007) and radiation
hydrodynamics simulations (Nelson et al.~2000; Cai et al.~2006, 2008; Boley et
al.~2006, 2007b; Stamatellos \& Whitworth~2008a; Boley \&
Durisen 2008) find that disk fragmentation inside $r\sim$ tens of AU is unlikely because
regions with high cooling rates and low $Q$ are rare (cf, most recently, 
Boss 2008 and Mayer et al.~2007). 
When $Q$ is low, the gas is cool, so cooling times are typically long.  When cooling
times are short, the gas is hot, so $Q$ is typically high.  In contrast, Boss (2004) argues that convection can increase the cooling rates enough to cause fragmentation.  Likewise, Mayer et al.~(2007) report convection-induced fragmentation in their simulations when the mean molecular weight is increased from 2.4 to 2.7\footnote{Boley et
al.~(2006) simulated a disk with a $\mu\sim2.7$; they did
not see fragmentation.}. Boley et al.~(2006) and Boley \& Durisen (2008) argue in return that the reported convective motions are likely shock bores along the spiral shocks (Boley \& Durisen 2006) or an artifact of suddenly changing the mean molecular weight.   Moreover, Rafikov (2007)
and Boley et al.~(2006, 2007b) find that convection does not decrease cooling times enough to trigger fragmentation because sustained convection is regulated by the entropy gradient and energy must be radiated away at the disk's surface.    Core accretion seems to be the only option for gas giant formation
inside $r\sim 100$ AU.  However, the recently-discovered substellar companions with wide semi-major axes $a\sim 100$ AU and
masses $M\sim$ few to tens $M_{\rm J}$ (e.g., Luhman et al.~2006; Lafreniere et al.~2008; Kalas et al.~2008) 
challenge the interpretation that core accretion is the sole formation mechanism for gas giants. The susceptibility of disks to fragmentation at large $r$ remains an open question.  Stamatellos et al. (2007; hereafter SHW2007) and Stamatellos \& Whitworth (2009; hereafter SW2009) find that extended, massive disks can fragment  under realistic conditions (cf Boss 2006); but, their results are based on running simulations with highly unstable initial conditions, which as argued here, can significantly change the outcome of disk evolution.  

Consider a disk where the sound speed becomes constant with radius
for large $r$. For
$\Sigma\sim r^{-q}$, $Q\sim r^{q}\Omega\sim r^{q-1.5}$.  As long as $q<1.5$,
the stabilizing shear contribution to $Q$ falls off faster than the destabilizing self-gravity
contribution, and the disk can become
susceptible to GIs. Now consider the radiative cooling time for some 
fluid element $t_{\rm
rad}=\epsilon/\vert\nabla\cdot{\bf \it F}\vert$, where $\epsilon$ is the internal
energy density of the gas.  Using the free-streaming limit for the divergence of the flux,  where $\vert\nabla\cdot {\bf\it F}\vert=4 \rho \kappa_p \sigma T^4$ for low optical depth $\tau=\int_{-\infty}^{\infty}\rho\kappa_P dz$, the cooling time can be approximated by   
$t_{\rm rad}\sim \Sigma c_s^2/(4\tau\sigma T^4)$, where $\rho$ is the mass density  at temperature $T$ and $\kappa_P$ is
the Planck mean opacity.  For $\tau\ll 1$, radiative
cooling becomes extremely inefficient.  Because cooling in the optically thick disk is also inefficient, the $\tau\sim 1$ transition region
may be the most conducive regime for disk fragmentation 
 (e.g., Rafikov 2005). The cooling time in a disk with $\Sigma\sim 1$ g cm$^{-2}$, $\tau\sim
0.1$, and $T\sim 10$ K, is  $t_{\rm rad}\sim 600\ {\rm yr}$, which
is less than the orbital period around a 1 $M_{\odot}$ star at $r\sim 100$ AU.

What if the $\tau\sim 1$ and $Q\sim1$ regions do not coincide?  Mass loading, i.e., mass accretion onto the disk,  is crucial for
driving fragmentation at $r>100$ AU.  Numerical fragmentation studies have focused mainly on thermal energy balance arguments. However, there are additional considerations when evaluating the stability of a disk against fragmentation. The rate that $Q$ changes in a local region of a disk is
\begin{equation}
\frac{d\ln Q}{d t}= \frac{1}{2}\frac{d \ln c_s^2}{d t}+ \frac{d \ln \kappa}{d t} - \frac{d\ln \Sigma}{d t}.
\end{equation}
When GIs are in a self-regulating phase, the effective Shakura \& Sunyeaev (1973) $\alpha$ tends toward a few hundredths
or less (e.g., Gammie 2001; Lodato \& Rice~2004; Mejia et al.~2005; Boley et al.~2006), so the timescale for $\Sigma$ and $\kappa$ to change  due to mass transport should be much larger than the local dynamical time. For
fragmentation criteria studies of {\it isolated disks}, the thermal energy term is the
most relevant.  In order to avoid fragmentation, the local cooling time for the gas
$t_{\rm cool}>f(\gamma) t_d$, where $t_d$  is the local gas orbital period and $f(\gamma)$ is some function of the adiabatic index of order unity.  In the local model formalism, the $t_{\rm cool}$ criterion is associated with a  critical effective $\alpha_c\sim0.06$ (Gammie 2001; Rice et al.~2005).  This $\alpha$-limit indicates the point at which local energy dissipation by GIs cannot balance cooling, and as a result, the disk fragments. 

 In addition to energy transport, $\alpha_c$ can be related to a critical mass flux $\dot{M}_c\approx 3\pi\alpha_c c_s^2\Sigma/\Omega$.  Substituting $Q$ into this relation gives $\dot{M}_c\approx 3\alpha_c c_s^3/(G Q)$ (see also Goodman 2003; Matzner \& Levin 2005). At large $r$, where the temperature of the disk approaches the envelope temperature, the ratio $\dot{M}_c/\dot{M}_e\sim 3\alpha_c/Q$, where $\dot{M}_e\sim c_s^3/G$ is the mass accretion onto the disk from the envelope.  For an envelope temperature of $T=30$ K, $\dot{M}_e\sim 2\times 10^{-5} M_{\odot}\rm~yr^{-1}$.  Because $\alpha_c\sim 0.06$, mass loading can operate at multiple times the maximum local transport flux.  {\it Episodes of non-local transport and/or disk fragmentation should occur for realistic envelope accretion rates}.   

I present results from
radiative hydrodynamics simulations that show mass loading can drive a disk toward fragmentation at large $r$.  The fragments form clumps with masses greater than a few $M_J$.   Throughout this Letter, I refer to the inner disk as the region
inside $r\sim 100$ AU and the extended disk as the region outside this
boundary.  This transition radius is illustrative, but represents a reasonable estimate for where fragmentation becomes possible (e.g, Matzner \& Levin 2005).

\section{Methodology}

I use CHYMERA (Boley 2007) to model the formation of massive extended disks. CHYMERA is an Eulerian
code that solves for the equations of hydrodynamics with self-gravity on an
evenly spaced, cylindrical grid. The rotational states of molecular hydrogen
are taken into account (Boley et al.~2007a), and the radiative transfer scheme has passed a series of analytic tests.  The radiative transfer algorithm and test results are presented in Boley et al.~(2007b).
D'Alessio et al.~(2001) opacity tables are used for
calculating optical depths, with a maximum grain radius $a_{\rm max}= 1\mu m$.
In the following simulations, an equilibrium ortho:parahydrogen ratio (O:P) is used. The gas is assumed to have a mixture of 0.73, 0.25, and 0.02 for hydrogen, helium, and metals, respectively.

Preliminary simulations using the CHYMERA radiative transfer scheme show that the radiative cooling time is shorter than the local orbital time and that the
disk is effectively isothermal, in agreement with the estimate in \S 1.  For modeling spiral waves, the normal radiative transfer scheme is stable in these simulations because the gas temperature is never far from the incident irradiation temperature.  However, for some preliminary simulations,
the cooling algorithm became numerically unstable immediately {\it after} clump formation due to the sudden increase in temperature, creating highly disparate radiative transfer and hydrodynamics timescales.  In order to 
study fragmentation at large $r$, the following radiative cooling algorithm is used for the simulations presented here.
The divergence of the flux
\begin{equation}
\nabla \cdot F = -(A/V) \sigma(T^4-T_{\rm irr}^4) f_{\rm \tau}^{-1},
\end{equation}
where $A/V$ is the cell area-to-volume ratio, $T_{\rm irr}$ is the incident irradiation on the disk at $r$, and $f_{\rm \tau}=\Delta \tau + 1/\Delta \tau$.  The local optical depth across a cell is calculated by 
$\Delta \tau = \rho (\kappa_{\rm Rosseland}  (1-\exp[-2\Delta \tau_{\rm Planck}])+\kappa_{\rm Planck} \exp[-2\Delta \tau_{\rm Planck}]) V^{1/3}$, where $\Delta \tau_{\rm Planck}$ is an initial estimate using the Planck mean opacity.  This cooling approximation goes to the free-streaming limit for small $\Delta \tau$ and to zero for large $\Delta \tau$.  In order to ensure that the algorithm is stable for large hydrodyamic time steps, the divergence of the flux is adjusted such that 
\begin{eqnarray}\nabla \cdot F^{\rm adjusted} &= &\nabla\cdot F \exp[-(\Delta t_{\rm hydo}/\Delta t_{\rm rad})^2]\\ \nonumber
 & +& \rho(e_{\rm equil}-e)/\Delta t_{\rm hydro} ( 1- \exp[-(\Delta t_{\rm hydro}/\Delta t_{\rm rad})^2]).
\end{eqnarray}
Here, $e$ is the specific internal energy of the gas, $e_{\rm equil}$ is the internal energy of the gas if it were at $T_{\rm irr}$,  $\Delta t_{\rm hydro}$ is the Courant time step, and $\Delta t_{\rm rad} = \rho e/\vert \nabla \cdot F\vert $  is the radiative timescale.  The irradiation temperature for these simulations is set to 30 K for all $r$.

The central protostar's position is integrated self-consistently with a softened potential, where $\Phi_{\rm star}=G M_{\rm star}/ (\vert {\bf r }-{\bf r'}\vert ^2 + s^2)^{1/2}$ for softening parameter $s$.  In order to treat the force on the star, the mass in each cell is treated as a point mass at the cell's center, with the same softening parameter used for the star.   The star's position is integrated from step $i$ to $i+1$ by the following: ${\bf v}^i = {\bf v}^{i-1/2} + 0.5 {\bf a}^i \Delta t^{i-1}$, ${\bf v}^{i+1/2} = {\bf v}^{i} + 0.5 {\bf a}^i \Delta t^{i}$, and ${\bf x}^{i+1} = {\bf x}^{i} + {\bf v}^{i+1/2} \Delta t^{i}$.  This algorithm is sufficient for maintaining, on average, the system's center of mass at the grid center.

Each disk is evolved on an $r,~\phi~,z=256,~512,~64$ grid, with a spatial resolution of $\Delta r$, $r \Delta \phi$, and $\Delta z$ = 2, $2 \pi r/512$, 2 AU.  Mirror symmetry is assumed about the midplane, and the outer grid boundaries are outflow boundaries.  There is an outflow boundary near the star, but negligible mass passes through it in these simulations.  Mass is added  to the grid near the top of the $z$ outflow boundary between $r=60$ and 300 AU.  The added mass is given a density profile of $\rho \sim r^{-p}$, a specific angular momentum $(G M_{\rm star}r^3/(r^2+s^2))^{1/2}$, an initial $v_r=0$, and an initial $v_z = -(2G M_{\rm star}/r)^{1/2}$.    Upon reaching a disk mass of 0.1 $M_{\odot}$, a random density perturbation is imposed with a maximum variation of $\pm10$\%.  For all simulations, the disk is stable against GIs when the noise is added.   For two of the simulations (see below) a softening of $s=20$ AU was applied.  This softening causes an error in the epicyclic frequency of just under 10\% at $r\sim 50$ AU when compared with the Keplerian frequency and 3\% at $r\sim100$ AU.

\section{Simulations and Results}

Four simulations (SIMA, SIMB, SIMC, and SIMD) are shown in Figure 1.  All disks except for SIMC fragment, and all disks evolve isothermally except in very high-density regions.   Because the vertically integrated midplane $\tau$ never becomes larger than a few, except in clumps, the above radiative transfer approximation  is reasonable.  Clumps reach temperatures in excess of 100 K. 

(1) SIMA: The protostar is set to 0.3 $M_{\odot}$.  The initial mass loading ${\dot{M}_d}\sim10^{-4} M_{\odot}\rm~yr^{-1}$ until the disk mass $M_d= 0.1 M_{\odot}$, after which ${\dot{M}_d}$ is reduced to $10^{-5} M_{\odot}\rm~yr^{-1}$.  The infall density profile is set to $p=1.5$. A softening $s=20$ AU is used because the accretion rate is less than what is used in the other simulations, requiring a longer evolution.  The simulation is evolved for 16400 yr  (3 $P_{200}$, orbital periods at $r\sim200$ AU), and reaches a disk mass of 0.21 $M_{\odot}$ before the simulation is stopped.  Dense spiral waves develop, and a condensation forms near $r\sim 70$ AU.  Fragmentation appears to be near wave corotation, in agreement with Durisen et al.~(2008).  By the end of the simulation, the clump has grown to 20 $M_J$ and is located at $r \sim 110$ AU.  As a cautionary check, a portion of this simulation was rerun with CHYMERA's normal radiative transfer algorithm. The disk behaved isothermally.

(2) SIMB: The protostar is set to 1 $M_{\odot}$, and $\dot{M}_d\sim10^{-4} M_{\odot}\rm~yr^{-1}$ for the duration of the simulation.  The simulation is evolved for 5000 yr ($\sim1.8$ $P_{200}$), and the disk grows to 0.52 $M_{\odot}$. The infall density profile $p=1.5$, and no softening is applied.  Mass loading drives $Q$ below unity outside $r\sim 100$ AU, and the disk fragments into 11 condensations by the time the simulation is stopped, ranging in mass from $\sim4$ to 14 $M_J$.

(3) SIMC: Similar to SIMB, but the mass accretion is halted when $M_d=0.33 M_{\odot}$.  At this mass, $Q$ in a region near $r\sim 90$ AU drops below unity.  This simulation investigates whether the disk can recover from the initial mass loading and avoid fragmentation.  When GIs did set in, they rapidly transported mass both away and toward the star, so a softening of $s=20$ AU was applied  and the inner disk boundary was moved from $r\sim 20$ AU to $r\sim 6$ AU.  The initial simulation without softening shows a similar behavior, but was not evolved beyond the initial burst of GIs.  The burst transports mass efficiently and $Q$ reaches a mass-weighted average $Q\sim 1.3$ between $r\sim 100$ and 200 AU.  This simulation is evolved for 9200 yr  ($\sim3.3$ $P_{200}$). Although a longer integration may be required to ascertain whether the disk will eventually fragment, the simulation does indicate that fragmentation is not guaranteed in the outer disk.

(4) SIMD: The protostar is set to 1 $M_{\odot}$, and the mass accretion rate is $10^{-4} M_{\odot}\rm~yr^{-1}$.  As in SIMC, the mass accretion is  halted once $M_d=0.33 M_{\odot}$.  The density profile for the infalling material $p=0.5$.  This choice for $p$ places the minimum $Q$ further out in the disk, such that $Q$ first drops below unity near $r \sim 140$ AU. Unlike SIMC, the spiral waves are unable to redistribute the mass enough to avoid fragmentation.  The disk forms a 6 $M_J$ clump at $r\sim 140$ AU, which is subsequently transported to just inside $r\sim 100$ AU and grows to 11 $M_J$ by the end of the simulation.  The disk is evolved for 8600 yr ($\sim 3$ $P_{200}$).

Wide semi-major axis, $a$, gas giants (W$a$GGs) and at least one brown dwarf (BD) are formed in these simulations. All disks go through strong burst-like phases, consistent with the arguments in \S 1. Although a high $\dot{M}$ is used for SIMB, SIMC, and SIMD,  disk fragmentation is still expected for $\dot{M}\sim10^{-5} M_{\odot}\rm~yr^{-1}$, as seen in SIMA.  
The local Jeans length is resolved by at least four cells throughout the simulations (Truelove et al.~1997; Nelson 2006), where the local cell size is taken to be the geometric mean of the cell dimensions.  These results indicate that fragmentation at large disk radii should be common and that W$a$GGs can be explained by {\it in situ} formation.  These simulations demonstrate that the number or fragments and the fragment masses depend on how the disk is assembled.  Studies investigating the clump mass spectrum of extended, massive protoplanetary disks, e.g., SW2009, are very sensitive to initial conditions.

Although the fate of these clumps is unknown, the results of Mayer et al.~(2004) and SW2009 indicate that some condensations can survive.  Radiative feedback, which is neglected in the current radiative cooling algorithm, should limit clump growth.  Detailed simulations that can resolve the photosphere of a fragment are required to address this concern, which is beyond the resolution limit in these simulations.  

These disks are evolved through only one burst. Vorobyov \& Basu (2005; hereafter VB2005) found in their 2D simulations that several bursts can occur during the disk accretion phase.  Whether a system forms W$a$GGs/BDs and retains them may be dependent on the details of the {\it last} GI burst.  Because these simulations suggest that disks should be susceptible to fragmentation during their formation, they also support the possibility of a clump-driven outburst mechanism for FU Orionis objects (VB2005).  
The FU Orionis phenomenon is
characterized by a rapid increase in the optical brightness of a young T Tauri
object, typically 5 magnitudes over a few to tens of years. Accretion from the inner disk onto the star is estimated to be as high as $10^{-4} M_{\odot}~\rm yr^{-1}$ (Hartmann \& Kenyon 1996). 
Because FU Ori objects have
decay timescales of $\sim100$ yr, an entire minimum mass solar nebula ($\sim0.01~M_{\odot}$) can be accreted onto the
protostar during an event.  The best explanation for the
optical outburst is a thermal instability (TI) (Bell \& Lin 1994).  However, prodigious mass flux through $r\sim 0.1$ AU seems to be required to drive the TI. 
A tidally-disrupted W$a$GG could supply the inner disk with 0.01 $M_{\odot}$ mass, even when only the extended disk is gravitationally unstable. 
 Helled et al.~(2006) calculated the contraction time for a Jupiter-mass clump, and found that such a clump takes  $3\times 10^5$ yr to reach a central temperature of 2000 K, i.e., the temperature required to dissociate enough H$_2$ to lead to rapid collapse.  This suggests that a W$a$GG has sufficient time to be transported inside $r \sim 1$ AU before it reaches mean densities that are high enough to avoid tidal disruption, where the disruption radius $r_t\sim 8\times10^{-3}(\bar\rho/{\rm g~cm^{-3}})^{-1/3}(M/M_{\odot})^{1/3}$ AU.  The Helled et al.~model remains at a mean density of $\bar\rho\sim2\times10^{-7}$ g cm$^{-3}$ for $\sim 10^{5}$ yr.  This density corresponds to a disruption radius $r_t\sim 1$ AU for a 0.3 $M_{\odot}$ star.  An immediate observational signature of this mechanism would be a large shift in the radial velocity of the protostar just before an outburst.

\section{The Two Modes of Planet Formation}

Boss's (1997) advancement of the disk instability mechanism has spawned a decade's worth of work examining
GIs as a
formation mechanism for gas giant planets (see Durisen et al.'s~2007 review).  One of the principal reasons for
the mechanism remaining contested for so long is that, like
core accretion, the mechanism works
under the right conditions.    Rafikov (2005, 2007), Boley et al.~(2006, 2007b), Stamatellos \& Whitworth 2008, and
Boley \& Durisen (2008) argue that radiative cooling timescales are too long
for fragmentation out to $r\sim 40$ AU and that sustained convection does 
not cause fragmentation.  This is also consistent with Nelson et
al.'s~(2000) 2D radiative hydrodynamics simulations, who assumed a polytropic
vertical density structure in order to calculate a photosphere temperature.
Convection pushes the entropy gradient toward zero, so a
vertical polytropic density stratification assumes efficient convection.
Observationally, the planet-metallicity relationship (Valenti \& Fischer~2005)
indicates that planet formation
favors a high-metallicity environment, which is strong evidence for core accretion (cf Boss 2005). The estimated core masses of
Jupiter and Saturn (Saumon \& Guillot 2004; Militzer et al.~2008), along with the ice giants, support the core
accretion mechanism. This evidence suggests that the dominant formation mechanism for gas giants inside $r\sim 100$ AU is core accretion. 
    
In contrast to the conditions inside $r\sim 100$ AU, optical depths should
approach unity for a substantial $\Delta r$ in the extended disk, as occurs in these simulations.  Efficient radiative cooling, long orbital periods, and an equilibrium O:P ratio
combine to favor fragmentation in the extended disk. W$a$GG formation as a result of mass loading represents the first mode of planet formation, and takes place in the first $10^{5}$ yr of the disk's lifetime.  Core accretion, which can continue after the main disk formation phase, represents the slower, second mode of gas giant formation.  The survival of W$a$GGs will likely depend on the disk and mass accretion conditions during the last burst of GI activity.

Both core accretion and disk instability are able to form gas
giants under suitable conditions.  In the inner disk, conditions favor core
accretion, and in the extended disk, conditions favor disk instability.
Scattering and planet-disk interactions should wash out any strict desert between the two formation regimes,
but a bimodal population of gas giant planet semi-major axes should still be
present.  Because core accretion is expected to be less efficient in low-metallicity systems, the ratio of W$a$GGs to planets at small $r$ should increase with decreasing metallicity.

I would like to thank the referee for pointing out a complementary study by Rafikov (2009, astro-ph/0901.4739), which was posted during the review of this manuscript.  Rafikov's work discusses fragmentation conditions and the stability of disks against high $\dot{M}$'s, while the study presented here demonstrates that a disk can be  pushed  realistically toward fragmentation.

\acknowledgments{I thank R.~H.~Durisen, L.~Mayer, G.~Lake, R.~Teyssier, O.~Agertz,  D.~Stamatellos, and the referee for comments that improved this manuscript.  This research was supported by a Swiss Federal Grant and the University of Zurich.  The presented simulations were run on NASA Advanced Supercomputing facilities.  }

\newpage

\newcommand{\chondrites}{2005, in ASP Conf.~Ser.~341, Chondrites and the protoplanetary disk}

\begin{figure}[h*]
\begin{center}
\includegraphics[width=8cm]{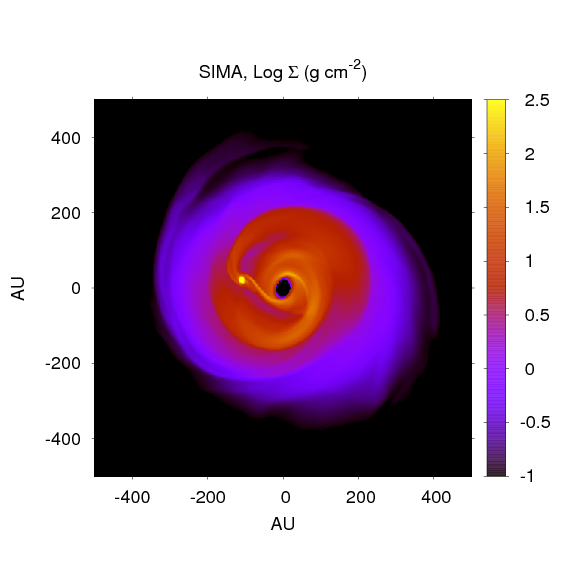}\includegraphics[width=8cm]{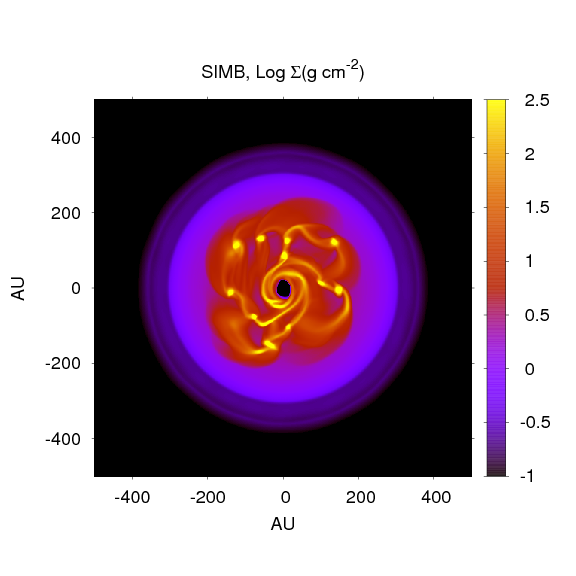}
\includegraphics[width=8cm]{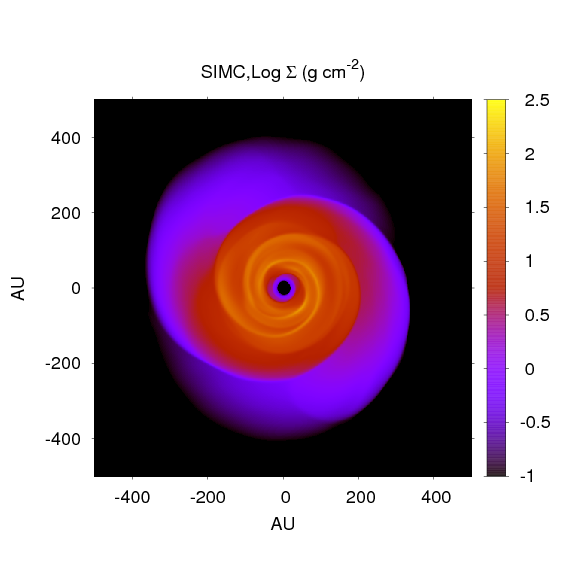}\includegraphics[width=8cm]{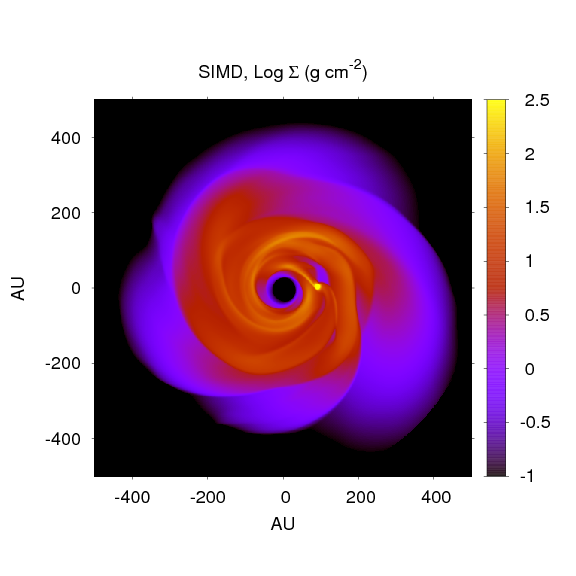}
\caption{Surface density snapshots at the end of each simulation.  Strong bursts of GI activity occur in each disk.  For three of the simulations, these bursts lead to fragmentation and clump formation.  Movies of these simulations can be viewed at http://www.aaroncboley.net  under the Movies tab. }
\label{default}
\end{center}
\end{figure}

\newpage

\end{document}